# Light-tunable optical cell manipulation via photoactive azobenzene-containing thin film bio-substrate.

Olivier Lefebvre #[1], Sandra Pinto #[1,2], Khalid Lahlil [3], Jacques Peretti[3], Claire Smadja [4], Clotilde Randriamampita [2], Mireille Lambert &[2] and Filippo Fabbri &[1] *

[1] Université Paris-Saclay, CNRS, C2N, 91120, Palaiseau, France

[2] Université de Paris, Institut Cochin, INSERM, CNRS, F-75014 PARIS, France

[3] Laboratoire de Physique de la Matière Condensée, Ecole Polytechnique / CNRS, Palaiseau, France

[4] Université Paris-Saclay, CNRS, Institut Galien Paris-Saclay, 92296, Châtenay-Malabry, France

* Corresponding author

# These authors contributed equally to this work

& co-senior authors

*KEYWORDS: Azobenzene, biomechanics, thin films, photoactive materials, tunable materials, smart materials*

**ABSTRACT:** In-vivo, real-time study of the local and collective cellular biomechanical responses requires the fine and selective manipulation of the cellular environment. One innovative pathway is the use of photoactive bio-substrates such as azobenzene-containing materials, which exhibit spectacular photomechanical properties, to optically trigger the local, mechanical stimulation of cells. Excited cells exhibit spectacular morphological modifications and area shrinkage, which are dependent on the illumination. This demonstrates the capabilities of photomechanically active substrates to study the phenomena resulting from the mechanical interaction of cells with their environment.

### 1. Introduction

Cells interact with a chemical and mechanical environment that influences their adaptive response and conditions the physiology and pathophysiology of the tissues. Much effort is being made to reconstruct simple systems in order to study the reciprocal influences of tissues and their environment [1-2], as well as to measure the consequences of environmental disturbances on cells. Current adopted technologies are magnetic micropillars [3], PDMS stretching [4], optical and magnetic clamps, microstructured polymers culture substrates with controlled rigidity [2,5] and rigidity gradients [6]. Despite this growing number of studies no technology allows in-situ, in-vivo and dynamic modifications of the substrate for a cellular or sub-cellular scale response to stress gradient change, which is relevant in many cell types [7].

The use of light as the control stimulus is a promising approach for cell stimulation, since it presents several advantages over the usual methods: non-contact operation (avoiding system perturbation or damage), remote actuation and sensing (compatible with restrictive environments and suitable for parallel addressing), high selectivity (wavelength, polarization, diffraction, etc.), local control down to submicronic scales. These features are particularly suitable for biological entities manipulation and analysis.

In this context, the photomechanical properties of photochromic materials such as azobenzene-containing systems offer a new way to optically achieve and control the mechanical stimulation of biological objects [8-13]. Azobenzene-type molecules undergo photoisomerization between TRANS and CIS isomers. These interconversion processes induce changes in the molecule physical properties, such as the dipole moment, which, for example in a thin film, results in the variation of surface properties such as wettability [14]. In addition, in the case of the so-called push-pull systems, such as the Disperse-Red One (DR1) derivative, the TRANS isomer is stable while the CIS isomer is metastable. Therefore, a cyclic conformation change is produced under illumination with a single wavelength. When incorporated in a polymer-like matrix, the repeated photoisomerization of the chromophores induces a spectacular mechanical response of the host material. First, the photoinduced modification of the viscoelastic behaviour of the matrix is observed. The mobility of molecules due to photoisomerization induces a reduction in the hardness and viscosity of the material, that is restored when the illumination ceases [15]. Second, a deformation of the material occurs under illumination due to efficient polarization-directed photo-induced mass migration processes. This was exploited for optical nanopatterning of polymer thin films and for photomechanical actuation of micro-nano structures [8-13]. These phenomena offer a new and unique way of creating substrates whose mechanical properties can be dynamically and locally controlled by light.

In this work, by using fluorescence microscopy techniques, we study the response of cells to the modification of their biomechanical environment induced by illuminating a DR1-containing polymer bio-substrate. We show that a cell shrinking can be obtained, whose amplitude depends on the duration of the photomechanical stimulus. This cellular response is not observed on a bare glass substrate under identical illumination conditions, thus confirming the fundamental role of the DR1-containing photoactive substrate. Both photoinduced deformation and viscoelastic properties change of the substrate can contribute to the observed cellular response.

## 2. Materials and methods

### Photoactive substrate

The photoactive substrate is an organic polymer film obtained by spin-coating a solution of PMMA-DR1 (Sigma Aldrich 570435) in dicholoromethane (25 mg/ mL) at 5000 rpm for 40 s at room temperature on a glass coverslip. The obtained thin film has a thickness of about 250 nm [12]. The volumic concentration in DR1 molecules is of the order of 1 molecule.$nm^{-3}$. The light absorption spectrum of the film, shown in Fig. 1(a), is peaked at 491 nm.

### Cell culture

The C2C12 mouse myogenic cell line was cultured at 37°C in 5% $CO_2$ in DMEM supplemented with 10% fetal bovine serum (FBS) [16]. After trypsinization, cells were deposited on a glass coverslip coated with PMMA-DR1 at $3x10^4$ cells/ml and cultured overnight.

### Actin staining

Before light stimulation of the substrate and observations, the actin cytoskeleton was stained by 45mn incubation with 6µM of SiR Actin (Cytoskeleton). Cells were washed 3 times with DMEM 10% FBS without phenol red and incubated in this medium. Coverslips were mounted in a Chamlide holder (GATACA Systems) for microscopic observations.

### Laser stimulation of azobenzene-containing polymer and wide-field microscopy

Samples are observed with an inverted iMIC *TILL Photonics* microscope provided by the Imag'IC facility (Institut Cochin, Paris). Laser stimulation was performed using multi-wavelength laser (including 491 nm - Cobolt Calypso 50 mW and 641 nm - Toptica iBREAM 100 mW). A schematic of the experimental setup is shown in Fig. 1(b). Two EMCCD cameras ANDOR Technology coupled with two lenses 1.5x (TuCam Andor Technology) are used to acquire images. Samples are observed with a x40 immersion objective throughout the experiment. The azobenzene substrate was excited with a 491 nm laser (3mW) for 4, 8 or 15 mn as described in the "Experimental results" section. The fluorescence image of the SiR Actin at 677 nm was performed with the same rate all along the illumination sequence and well after the 491 nm laser was switched off.

Experiments were performed with the help of the BioMecan'IC facility (Institut Cochin, Paris).

### Image analysis

For each image, we extract the area of the cells, by using the routine area measurement algorithm of FIJI software suite. We then calculate the averages of the measured cell areas for each illumination time, and we compute the statistical significance by two-way ANOVA with Tukey's multiple comparisons test.

### AFM setup

We used an AFM system (Bruker DI3100) in "Tapping mode" with a Si AFM tip (Nanosensor PPP-NCHR). We worked in air at room temperature with a low free amplitude (about 200 mV) to be soft enough not to damage or modify the measured surface. The tetrahedral tip with the sharper tip (tip radius less than 10 nm) was used for high lateral resolution. The nominal resonant frequency and the spring constant of the AFM cantilever were 330 kHz and 42 N/m.

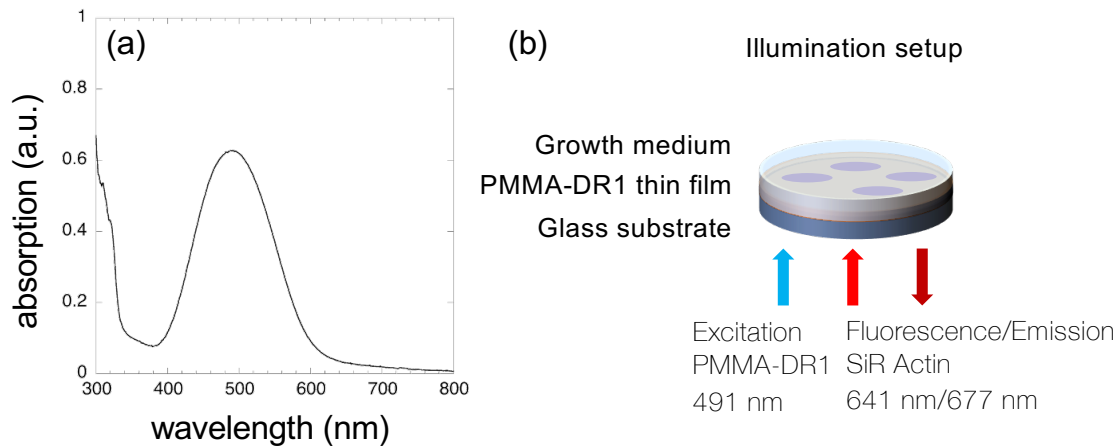

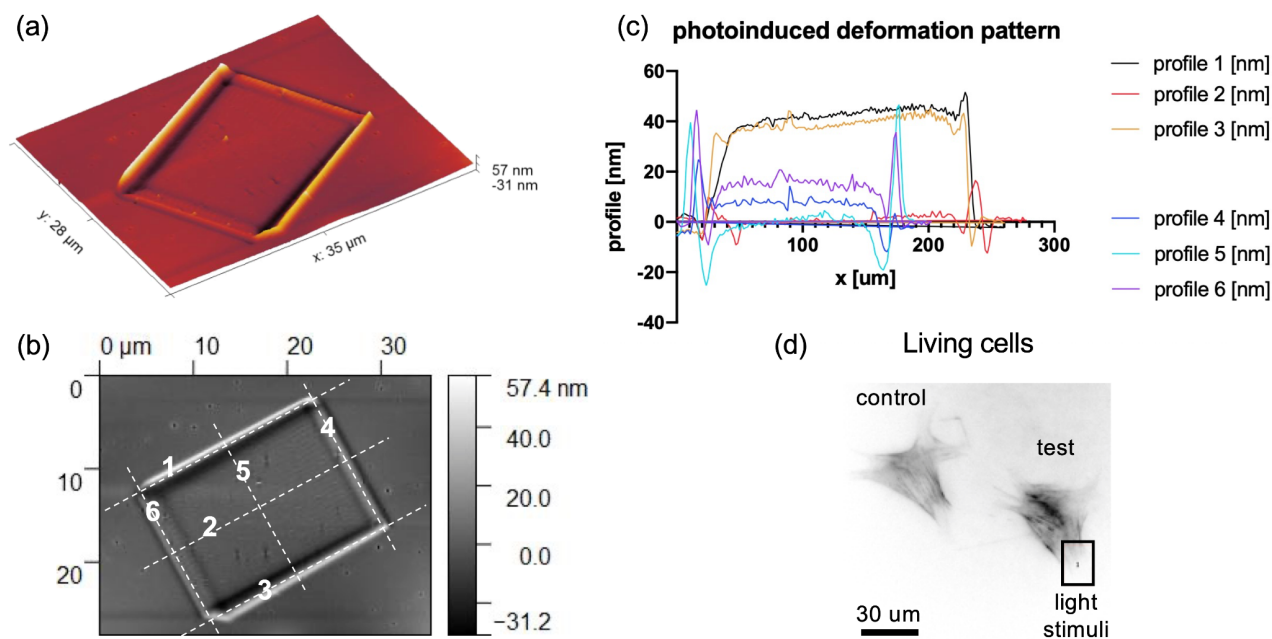

*Figure 2. (a-c) AFM topography measurement of the illuminated area and corresponding profiles; (d) Fluorescence actin imaging of C2C12 mouse cells, indicating the light stimuli zone.*

### 3. Experimental results

Two main photomechanical phenomena can be generally triggered in DR1-containing polymers: photo-induced deformation (DEF) and photoinduced modification of the material's viscoelastic properties (VISC), in particular of the hardness and of the viscosity. The DEF effect is mostly a permanent mechanical modification of the polymer matrix, which remains after the light stimulus is switched off. On the other hand, the VISC effects are not permanent, and the material's mechanical properties return to their original state once the illumination is switched off.

We optically trigger these mechanisms in the PMMA-DR1 layer in order to mechanically stimulate cells in-vitro, in specific localized areas and with specific time profiles.

In order to study the effect of each factor on the cells, we use a rectangular illumination pattern (wavelength: 491 nm, power density: 3 mW/mm2, illumination time : 4, 8, 15 min), that excites both the DEF and VISC effects on the edges, while it excites only the VISC in the surface of the rectangle (Fig. 2a), via the cyclic projection of an array of illumination lines: the excitation beam is continuously scanned over the rectangle area shown in Fig.2b. By Atomic Force Microscopy, we measure photo-induced deformations of a few tens of nm in height on the rectangle edges; no significant deformation is observed within the rectangle's surface (see profiles 2 and 5 in Fig. 2b-c). We obtain a permanent DEF effect on the edges, while we induce a transient VISC effect on the inner surface of the rectangle. The typical positioning of the photomechanical excitation pattern with respect to a living cell is shown in Fig. 2d. The black rectangle indicates the optically excited area on the cell's lamellipodia, where the motility is higher.

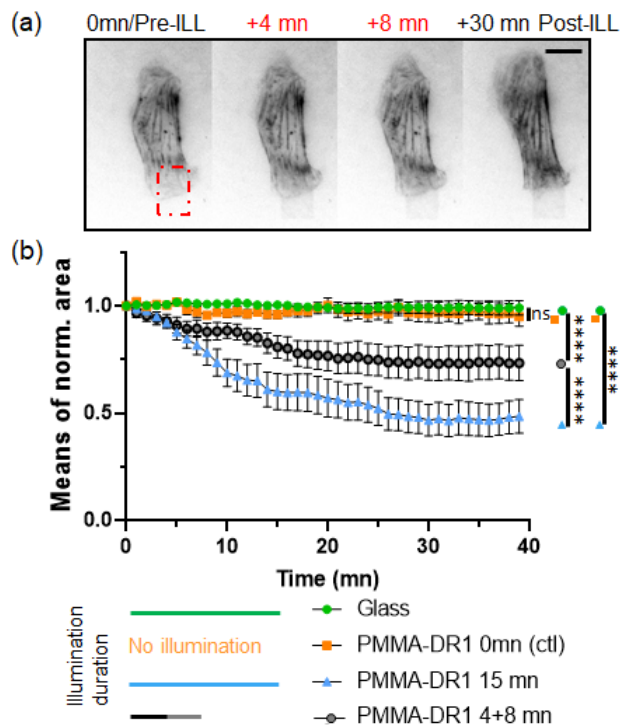

*Figure 3:* Cell excitation experiments on PMMA-DR1-coated glass substrate with illumination for 0, 4, 8 and 15 mn. Control experiments in the absence of illumination on the PMMA-DR1-coated glass slide (PMMA-DR1 0mn (ctl)) and with illumination for 15 mn on bare glass substrate are also presented. *(a)* Representative SiR Actin fluorescence images of C2C12 cells before illumination, during 8 mn of illumination and 30 mn after illumination is turned off. The red rectangle shows the light-stimulated area. The scale bar is 10 µm. *(b)* Variation versus time (from the illumination switch on) of the cells' area normalized to the average value measured over the first 5 mn before illumination. "ns" non statistically significant, **** p-value<0.0001 by two-way ANOVA with Tukey's multiple comparisons test. Several independent experiments were performed: PMMA-DR1 15mn, (n=14 cells); 4 & 8 mn (n=16 cells); 0mn (n=7 cells); Glass (n=6 cells). Pre-ILL = pre-illumination. All error bars are SEM.

Fig. 3 shows the observed evolution of the cellular responses under linearly polarized blue light stimuli (491 nm, 3 mW) of the photoactive PMMA-DR1 substrate, operated on the cells' lamellipodia, for three illumination times (4, 8 and 15 mn). For each experiment, the cells' behavior is monitored by optical and fluorescence microscopy, before, during and after blue laser illumination. SiR Actin fluorescence images are captured every minute from which is measured the evolution of the cell's area. Average evolution of area parameter over time was normalized relatively to the mean area value measured 5 mn before illumination is switched on.

Fig. 3a shows the images of a typical cellular response observed on light-stimulated PMMA-DR1 layer. On the image recorded before the illumination is switched on (0mn), the red rectangle indicates the illuminated area. After 8 min illumination, a shrinkage of the cells' lamellipodium is observed at the bottom of the image near the substrate illumination area. Note that the cell keeps the capacity for spreading a new lamellipodium at the top of the image (+30 mn post-illumination). The graph on Fig. 3b quantifies the responses of the cells for three illumination times and over the 40 mn of the experiment duration. For comparison, two control experiments where performed with cells deposited: i) on a glass substrate without PMMA-DR1 layer but illuminated for 15 min (Glass condition - green line) and ii) on the PMMA-DR1 coated substrate, but without light stimuli (PMMA-DR1 0mn ctl - orange line).

Our statistical analysis shows that the overall photoinduced area shrinking kinetics depends on the illumination duration. The efficiency of the cellular response increases with the illumination time: an illumination of 4 or 8 min (gray curve) is sufficient to stimulate a moderate but significant cellular response characterized by a modest area shrinkage of about 30% and slow dynamics. By increasing the illumination time to 15 min (blue curve), we obtain a larger cellular response with a 50% area shrinkage (p-value<0.0001 in comparison to "4mn + 8mn").

These observations clearly show the effect of the optical actuation of the photoactive PMMA-DR1 layer on the cell mechanical response, with respect to the control experiments, where no significant effect on the cell behavior is observed (p-value<0.0001).

Note that, during these experiments, cells continue their natural biological functions, which causes random area variations, which may be independent from the optical stimuli.

## 4. Discussion and conclusions

In our experiments, the cells' mechanical response is observed only in the presence of the PMMA-DR1 layer and subsequently to illumination in the absorption band of the photoactive substrate. It is thus clearly resulting from the photomechanical phenomena triggered in the PMMA-DR1 under light absorption. However, different photomechanical phenomena are at play in the PMMA-DR1 layer.

On the one hand, it is known that DR1-containing polymers may exhibit two photo-deformation mechanisms: a matrix photo-expansion and a mass migration process [17]. The matrix photoexpansion is moderate (a few percent of the total film thickness), homogeneous and fully completed within only a few seconds of illumination. It is thus very improbable that it plays a major role on the cell shrinkage. The matter migration process may be triggered by two different mechanisms: one related to the light polarization spatial distribution and the other one governed by the light intensity spatial distribution [12]. Although it is difficult to quantitatively describe the local distribution of light polarization and intensity, due to the influence of the liquid physiological environment and on the medium local fluid mechanics, the substrate photo-deformation observed on the edge of the illuminated area shows that at least one photo-deformation mechanism was activated. It is however difficult to determine the influence of this local deformation on the cells' response.

On the other hand, the photoinduced modification of the viscoelastic properties of the azobenzene-containing polymer may also play a role with respect to the cells' mechanical response. Indeed, the polymer photoinduced softening in the illuminated area may induce a change into the adhesion forces and the lamellipodia stability, hence inducing a mechanical response of the cell in order to improve mechanical stability. Moreover, the photoinduced change of the DR1 dipole moment may affect the surface hydrophilic/hydrophobic properties, inducing cellular mechanical reactions.

Finally, the kinetics of the cellular response, in the order of tens of minute, is slower than the typical photo-deformation kinetics (typically a few minutes) and the visco-elastic response (typically a few seconds) observed in DR1-containing polymer thin films [12, 15, 17]. This suggests that a complex mechanical cellular response is triggered in order to adapt to the new environment.

In conclusion, this work brings new evidence of the possibility of optically controlling the cellular environment by means of the light-stimulation of photoactive polymer bio-substrates. Further investigation will be needed to identify the mechanisms at play at the cell/substrate interface and their influence on the intracellular and intercellular dynamics.

## Acknowledgments

We acknowledge the help provided by Pierre Bourdoncle and Gabriel Le Goff (Imag'IC platform at Institut Cochin, Paris) and Christophe David (Centre de Nanosciences et Nanotechnologies, Université Paris-Saclay) for their important guidance of iMIC microscopy and AFM (Atomic Force Microscopy), respectively. We thank Heyem Neche for her help in some analytical AFM datas. We also acknowledge the BioMecan'IC platform at Institut Cochin for its fruitful discussions and irreplaceable contribution in biological experiments, and the RENATECH platform at C2N.This work was supported by the LABEX Lasips.

## References

[1] Engler, A.J., Sen, S., S., Sweeney, H.L., and Discher, D.E. Cell 126, 677-689 (2006).

[2] Ladoux, B., Anon, E., Lambert, M., Rabodzey, A., Hersen, and Al., Biophys. J. 98, 534–542 (2010).

[3] the Digabel, J., Biais, N., Fresnais, J., Berret, and Al., Lab. Chip 11, 2630-2636 (2011).

[4] Carpi, N., and Piel, M.,. J. Vis. Exp. JoVE (2014).

[5] Ganz, A., Lambert, M., Saez, A., and Al., Biol. Cell Auspices Eur. Cell Biol. Organ. 98, 721-730 (2006).

[6] Lo, C.M., Wang, H.B., Dembo, M., and Wang, Y.L. (2000). Biophys. J. 79, 144–152.

[7] Galvagni, F., Baldari, C.T., Oliviero, S., and Al., Cell Biol. 138, 419-433 (2012).

[8] P. Rochon, E. Batalla, and A. Natansohn, App. Phys. Lett. 66, 136 (1995)

[9] D. Y. Kim, S. K. Tripathy, L. Li, and J. Kumar, App. Phys. Lett. 66, 1166 (1995)

[10] F. Fabbri, Y. Lassailly, K. Lahlil, J. P. Boilot, and J. Peretti, App. Phys. Lett. 96, 081908 (2010)

[11] F. Fabbri, D. Garrot, Y. Lassailly, K. Lahlil, and Al., J. Phys. Chem. B 115, 1363 (2011)

[12] F. Fabbri, Y. Lassailly, S. Monaco, K. Lahil, J. P. Boilot, and J. Peretti, Phys. Rev. B 86, 115440 (2012)

[13] D. Vu, F. Fabbri et al., SPIE Proc. 9236, 923611 (2014)

[14] K. Ichimura, et al. Science 288, 1624 (2000)

[15] L. Sorelli, F. Fabbri, J. Frech-Baronet, A.D.Vu, et al., J. Mater. Chem. C, 3, 11055 (2015)

[16] D.Yaffe and O.Saxel, Nature 270, 725 (1977)

[17] D. Garrot, Y. Lassailly, K. Lahlil, J. P. Boilot, and J. Peretti, App. Phys. Lett. 94, 033303,(2009)